# Asymmetric metasurface photodetectors for single-shot quantitative phase imaging

Jianing Liu, Hao Wang, Yuyu Li, Lei Tian, and Roberto Paiella

Department of Electrical and Computer Engineering and Photonics Center,

Boston University, 8 Saint Mary's Street, Boston, MA 02215

ABSTRACT: The visualization of pure phase objects by wavefront sensing has important applications ranging from surface profiling to biomedical microscopy, and generally requires bulky and complicated setups involving optical spatial filtering, interferometry, or structured illumination. Here we introduce a new type of image sensors that are uniquely sensitive to the local direction of light propagation, based on standard photodetectors coated with a specially designed plasmonic metasurface that creates an asymmetric dependence of responsivity on angle of incidence around the surface normal. The metasurface design, fabrication, and angle-sensitive operation are demonstrated using a simple photoconductive detector platform. The measurement results, combined with computational imaging calculations, are then used to show that a standard camera or microscope based on these metasurface pixels can directly visualize phase objects without any additional optical elements, with state-of-the-art minimum detectable phase contrasts below 10 mrad. Furthermore, the combination of sensors with equal and opposite angular response on the same pixel array can be used to perform quantitative phase imaging in a single shot, with a customized reconstruction algorithm which is also developed in this work. By virtue of its system miniaturization and measurement simplicity, the phase imaging approach enabled by these devices is particularly significant for applications involving space-constrained and portable setups (such as point-of-care imaging and endoscopy) and measurements involving freely moving objects.

## 1. Introduction

Traditional image sensors can only capture the intensity distribution of the incident light, whereas all information associated with the phase profile is lost in the image acquisition process. While these devices are clearly adequate for basic imaging tasks, direct access to the wavefronts and local directions of light propagation would allow for more advanced imaging capabilities. One example of particular interest is the ability to visualize phase-only objects where light is transmitted or reflected without any appreciable intensity variations. Relevant application areas where this capability plays a prominent role include microscopy for label-free imaging of biological samples [1], surface profiling, and semiconductor inspection for detecting manufacturing defects [2]. Conventionally, phase imaging is achieved with rather complex and bulky setups, ranging from Zernike phase-contrast and differential-interference-contrast microscopy to quantitative techniques based on interferometry [1] or non-interferometric methods [3, 4]. More recently, newly developed free-space nanophotonics and flat-optics platforms have also been applied to the demonstration of similar phase imaging systems, with the potential advantage of more compact dimensions and enhanced design flexibility [5-12].

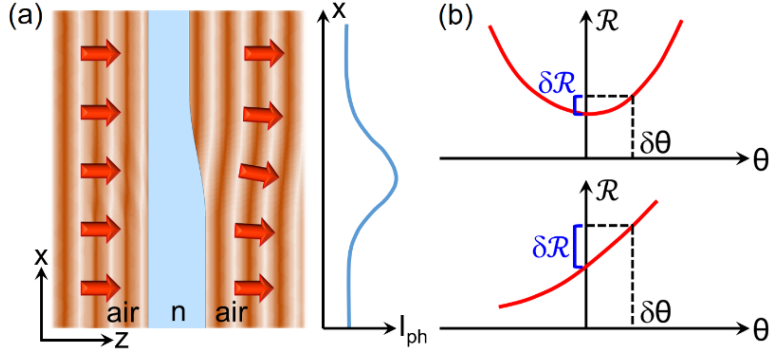

**Figure 1.** Phase contrast imaging with angle-sensitive photodetectors. (a) Left: wavefront distortion experienced by a plane wave after transmission through a transparent plate of variable thickness. Right: photocurrent signal $I_{ph}$ measured by an angle-sensitive photodetector at different locations across the transmitted wavefront. (b) Responsivity $\mathcal{R}$ versus angle of incidence $\theta$ for a generic device with symmetric (top) and asymmetric (bottom) angular response. In the limit of small deflection $\delta\theta$, the asymmetric device provides a larger change in responsivity $\delta\mathcal{R}$, leading to increased image contrast.

In this work, we report the development of image sensors that can measure the phase gradient of the incident optical field directly with the simplest possible setup, i.e., a standard camera or microscope without any external optical elements other than the imaging lenses. These devices consist of photodetectors individually coated with an integrated plasmonic metasurface that introduces a sharp dependence of responsivity $\mathcal{R}$ on illumination angle $\theta$ near normal incidence. The resulting wavefront sensing ability is illustrated schematically in Fig. 1(a), where a plane wave of field amplitude $U_{in}(z) = U_0 e^{ikz}$ is incident on a transparent object that introduces a position-dependent transmission phase shift $\varphi(x)$. Correspondingly, the direction of propagation of the transmitted wave $U_{tr}(x,z) = U_0 e^{i[kz+\varphi(x)]}$ is tilted to approximately $\hat{\mathbf{x}}\frac{d\varphi(x)}{dx} + \hat{\mathbf{z}}k$, i.e., by a position-dependent angle $\theta(x) \approx \frac{1}{k}\frac{d\varphi(x)}{dx}$. If the transmitted light is detected with an array of angle-sensitive photodetectors, the photocurrent signals $I_{ph}$ produced by different pixels at different x locations will therefore vary with the local phase gradient $\frac{d\varphi(x)}{dx}$ of the object. It also follows from this discussion that the contrast of the resulting image is ultimately limited by the photodetector responsivity slope $d\mathcal{R}/d\theta$ in the limit of small $\theta$ [see Fig. 1(b)]. As a result, devices with an asymmetric angular response (where $\mathcal{R}$ is linearly proportional to $\theta$ in the small-$\theta$ limit) are preferable for this application compared to symmetric devices (where $d\mathcal{R}/d\theta$ vanishes for $\theta = 0$). Additionally, an asymmetric response also allows for the unambiguous determination of the sign of the angular deflection.

In our directional image sensors, this desired angular asymmetry is produced by an array of Au nanostripes that selectively couple light incident at a target detection angle (slightly offset from normal incidence) into surface plasmon polaritons (SPPs) guided by an underlying metal film [Fig. 2(a)]. The excited SPPs are then scattered into the supporting photodetector active layer by a set of slits perforated through the metal film on one side of the nanostripe array. Light incident along any other direction is instead simply reflected or diffracted back. Devices based on a similar concept, with responsivity peaked at geometrically tunable angles over an ultrawide field of view of ~150°, have been reported recently to enable flat lensless compound-eye vision [13]. The same devices can also be used to perform optical spatial filtering with coherent transfer function determined by their angular dependent responsivity $\mathcal{R}(\theta)$, as shown by detailed theoretical modeling for representative symmetric structures in ref. 14. Alternative device configurations for angle-sensitive vision that have been demonstrated previously include the use of lenslet arrays [15], stacked gratings based on the Talbot effect [16], and micro-apertures across adjacent pixels [17]. For phase imaging applications, the key advantage of the configuration of Fig. 2(a) is the ability to be designed with particularly sharp asymmetric responsivity peaks of large slope $d\mathcal{R}/d\theta$. To demonstrate the resulting wavefront sensing capabilities, here we have developed a tailor made device for this application, measured its angle-dependent responsivity, and then used the experimental data in conjunction with computational imaging techniques to evaluate the phase contrast images produced by full pixel arrays of these sensors. Our results show that a minimum detectable phase contrast as small as 8 mrad can be achieved, highlighting the promise of these angle-sensitive photodetectors to substantially miniaturize and simplify phase imaging systems while still providing state-of-the-art sensitivity.

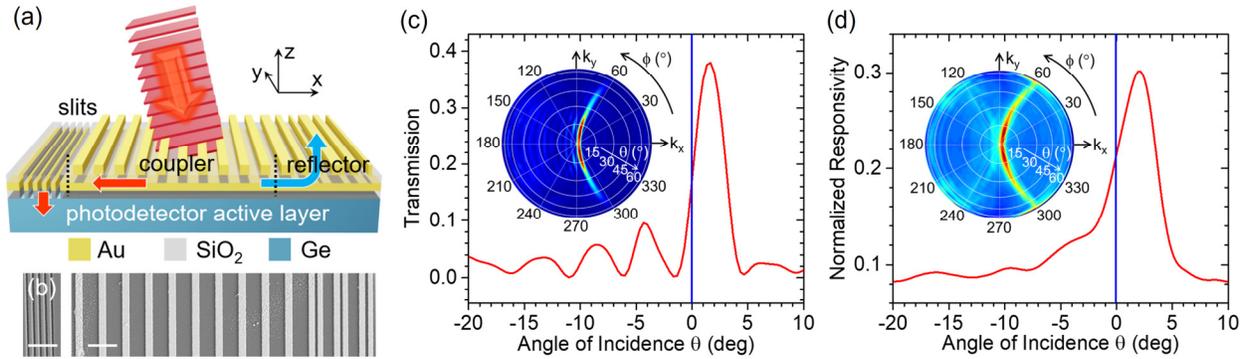

**Figure 2.** Asymmetric metasurface photodetectors. (a) Schematic device structure and principle of operation. (b) Top-view SEM images of an experimental sample, showing the slits (left image) and nanostripes (right). The scale bars are 2 μm. In this device, the metal film nominally consists of 5 nm of Ti and 100 nm of Au, the two $SiO_2$ layers have a thickness of 60 nm, and each grating line consists of 5 nm of Ti and 50 nm of Au with a width of 440 nm. The grating coupler contains 10 lines with a period $\Lambda = 1432$ nm. The slit section comprises 5 slits with 200-nm width and 400-nm center-to-center spacing. The reflector design is described in the Supplementary Material. (c) Inset: calculated transmission coefficient through the metasurface of this device for *p*-polarized incident light at $\lambda = 1550$ nm versus polar $\theta$ and azimuthal $\phi$ illumination angles. Main plot: horizontal line cut of the color map. (d) Inset: measured angular dependence of the responsivity of the same device, normalized to the normal-incidence responsivity of an identical photodetector without any metasurface. Main plot: horizontal line cut of the color map. The vertical blue lines in (c) and (d) indicate normal incidence.

The phase measurement carried out by these devices is conceptually similar to the differential phase contrast (DPC) approach, in which a reciprocal-space asymmetry is introduced in the sample illumination [4, 18, 19], in the pupil plane [20], or by split detectors in a scanning microscope [21], to convert phase gradients into intensity variations. This approach has been employed for *quantitative* phase reconstruction by sequentially recording one or multiple pairs of DPC images with mirrored asymmetric illumination [4, 18, 19]. The two images in each intensity pair are subtracted from each other to remove the unknown background, and the process of phase differentiation is then digitally inverted by a deconvolution algorithm. As shown in the following, the same idea can be implemented with an array of asymmetric angle-sensitive photodetectors where alternating pixels feature equal and opposite responsivity functions $\mathcal{R}_+(\theta) = \mathcal{R}_-(-\theta)$. With this configuration, the two mirrored DPC images required for background subtraction are acquired simultaneously (i.e., in a single shot) by the two types of pixels. Correspondingly, the overall measurement can be significantly simplified compared to previous quantitative DPC setups, because it does not require any specialized time-modulated directional sources [4, 18, 19] or beam scanning [21]. As a result, this approach is particularly promising for applications where space and time are highly constrained, such as point-of-care and in vivo microscopy, endoscopy, and imaging of freely moving objects.

## 2. Results and discussion

In the device architecture of Figs. 2(a) and 2(b), the illumination window of a photodetector is coated with a $SiO_2/Au/SiO_2$ stack. A periodic array of Au nanostripes (grating coupler) is then introduced over the top $SiO_2$ layer, surrounded on one side by a set of subwavelength slits perforated through the stack and on the other side by a short section of Au nanostripes of different widths (reflector). The Au film has sufficiently large thickness (100 nm) to block any incident light from being transmitted directly into the device active layer. As a result, photodetection can only take place through an angle-sensitive plasmon-assisted process where SPPs on the top surface of the metal film are initially excited via diffraction of the incident light by the grating coupler.

This process is governed by the Bragg condition $\sin\theta/\lambda \pm 1/\Lambda = \pm n_{SPP}/\lambda$, where $\lambda$ is the incident wavelength, $\Lambda$ is the grating period, $n_{SPP}$ is the SPP effective index, and the plus and minus signs correspond to SPPs propagating along the positive and negative x directions, respectively. Backward traveling SPPs eventually reach the slit section, where they are preferentially scattered into the photodetector active layer, similar to the phenomenon of extraordinary optical transmission through sub-wavelength apertures in metal films [22]. A photocurrent signal is then detected proportional to the SPP field intensity at the slit locations. In contrast, forward traveling SPPs eventually arrive at the reflector, which is designed to scatter them back into radiation propagating away from the device into the air above. Briefly, the nanostripe widths in this reflector section are selected to produce a linear scattering phase profile for the incoming SPPs (and therefore suppress all diffraction channels except for the -1 order) based on the notion of gap-plasmon metasurfaces [23, 24]. With this arrangement, all forward traveling SPPs can be scattered away from the device surface within the smallest possible area (see Supplementary Material for more details). Altogether, the composite metasurface comprising the metal film, grating, slits, and reflector therefore behaves like an angle-selective filter for the light transmitted into, and ultimately absorbed by, the photodetector. The required asymmetric angular response for quantitative phase imaging is enabled by the aforementioned diverging action of the slits and reflector on oppositely traveling SPPs.

The specific device developed in this work features a narrow responsivity peak $\mathcal{R}(\theta)$ centered at $\theta \approx 2°$, i.e., only slightly offset from normal incidence to maximize the slope $d\mathcal{R}/d\theta$ at $\theta = 0$. The key geometrical parameters, listed in the caption of Fig. 2, were optimized via finite difference time domain (FDTD) simulations. Because of the diffractive nature of the device operating principle, the angular peak position is sensitive to the incident wavelength, and operation near $\lambda$ = 1550 nm is considered throughout this work. The resulting phase imaging system is therefore primarily intended for monochromatic (i.e., laser light) illumination, although high spatial coherence is not needed (unlike typical interferometric setups, which correspondingly often suffer from speckle artifacts [1, 3, 4]). The grating-coupler nanostripe width w and period $\Lambda$ are 440 nm and 1432 nm, respectively, selected to produce efficient excitation of SPPs by light incident at the desired angle of peak detection (~2°) according to the Bragg condition. The number of nanostripes in the grating is 10, selected to minimize the angular width of the responsivity peak (based on the interplay between SPP propagation losses and diffraction effects), while at the same time maintaining a reasonably small pixel size (21.8 µm, including the slits and reflector section).

Figure 2(c) presents simulation results for the p-polarized power transmission coefficient of the optimized metasurface as a function of polar $\theta$ and azimuthal $\phi$ angles of incidence. The figure inset shows the full angular response across the entire hemisphere, obtained from a three-dimensional FDTD simulation based on the principle of reciprocity (see Methods). The main plot of the same figure shows the horizontal line cut of the color map (i.e., transmission versus $\theta$ for $\phi$ = 0). These simulation results reveal a narrow angular region of high transmission adjacent to normal incidence, with a characteristic C shape determined by the Bragg condition for the excitation of SPPs traveling along different directions. By design, the low-angle tail of the transmission peak is centered around $\theta$ = 0 (vertical blue line in the main plot). The maximum transmission coefficient (at $\theta$ = 1.6°) is over 38%, indicating that the transmission penalty introduced by the metasurface is reasonably small. Similar calculations for s-polarized incident light show negligible transmission at all angles, consistent with the longitudinal nature of SPP modes. As a result, these devices require polarized illumination for maximum detection efficiency.

If the metasurface just described is fabricated on the illumination window of an image sensor, the device responsivity can be expected to vary with angles of incidence exactly as in the color map of Fig. 2(c), regardless of the photodetector operating principle. Here, for convenience, we employ a Ge metal-semiconductor-metal (MSM) photoconductor, which simply consists of two Au contacts deposited on the top surface of a Ge substrate. The metasurface is then introduced in the space between the two electrodes with a multi-step fabrication process involving various thin-film deposition techniques and electron-beam lithography (see Methods). Figure 2(b) shows scanning electron microscopy (SEM) images of an experimental sample, highlighting the slits, grating, and reflector section. The completed device was characterized with angle-resolved photocurrent measurements under polarized laser light illumination. The incident wavelength $\lambda$ was adjusted to optimize the position of the responsivity peak relative to normal incidence for maximum $d\mathcal{R}/d\theta$ at $\theta = 0$. All the experimental results presented below were measured with $\lambda = 1610$ nm, about 4% larger than the design value of 1550 nm. This rather small discrepancy is ascribed to similarly small deviations of the sample geometrical parameters from their target values (for example, the thickness of the $SiO_2$ spacer layer above the Au film, which affects the SPP effective index $n_{SPP}$).

With this adjustment, the measurement results are in good agreement with the design simulations. As shown in Fig. 2(d), the measured responsivity peak is centered at 2.2° with a full width at half maximum (FWHM) of 5.5°, reasonably close to the calculated values of 1.6° and 3.0°, respectively, from Fig. 2(c). The vertical axis in Fig. 2(d) is normalized to the responsivity of an otherwise identical reference sample without any metasurface (see Supplementary Material). Correspondingly, a peak value of about 30% is obtained, again in reasonable agreement with the design simulations of the metasurface transmission. The smaller peak height and larger FWHM observed in the experimental data likely originate from residual roughness in the Au film, which decreases the SPP propagation length and thus reduces the fraction of SPPs captured by the slits. The inset of Fig. 2(d) also shows a weak signature of photocurrent measured through the excitation of forward traveling SPPs (faint C-shaped feature in the left half of the color map), which is attributed to a small misalignment of the slits relative to the grating section. However, this unintended photodetection channel does not significantly degrade the angular response asymmetry near normal incidence, as can be clearly seen in the line cut of the same figure.

Next, we consider an image sensor array based on the devices of Fig. 2 and evaluate its phase contrast imaging capabilities. To that purpose, we employ the frequency-domain model developed in ref. 14 to substantiate the use of similar plasmonic directional photodetectors for optical spatial filtering. The key conclusion of this model is that these devices sample the incident field distribution at their slit locations ($\mathbf{r} = \mathbf{r}_{sl}^n$ for the $n^{th}$ pixel in the sensor array), according to the coherent transfer function

$$t(\mathbf{k}) \equiv \frac{E_{SPP}(\mathbf{k})}{E_{in}(\mathbf{k})} \propto e^{-i\alpha k_x}\sqrt{\mathcal{R}(\mathbf{k})}. \tag{1}$$

In this equation, $\mathbf{k} = (2\pi/\lambda)(\hat{\mathbf{x}}\cos\phi + \hat{\mathbf{y}}\sin\phi)\sin\theta$ is the in-plane wavevector (with $\hat{\mathbf{x}}$ perpendicular to the slits and nanostripes), $E_{in}(\mathbf{k})$ and $E_{SPP}(\mathbf{k})$ are the spatial Fourier transforms of the incident and SPP fields on the sensor array $E_{in}(\mathbf{r}_{sl}^n)$ and $E_{SPP}(\mathbf{r}_{sl}^n)$, respectively, and $\mathcal{R}(\mathbf{k})$ is the angle-dependent responsivity. Finally, the phase slope $\alpha$ is approximately equal to the distance between the slits and the pixel center, depending on the SPP propagation losses [14]. The exact value of this parameter has actually no observable effect on the recorded images [by the shifting property of Fourier transforms, the phase factor $e^{-i\alpha k_x}$ in $t(\mathbf{k})$ simply corresponds to a uniform displacement in real space by the amount $\alpha$ in the –x direction]. In the following, we use $\alpha = 8$ μm, computed via FDTD simulations for the present device (see Supplementary Material).

For the responsivity function $\mathcal{R}(\mathbf{k})$ in eq. (1), we use the experimental data shown in the inset of Fig. 2(d). Finally, $E_{in}(\mathbf{k})$ can be related to the Fourier transform of the object in the field of view $E_{obj}(\mathbf{k})$ according to $E_{in}(\mathbf{k}) = t_{lens}(\mathbf{k})E_{obj}(\mathbf{k})$, where $t_{lens}(\mathbf{k})$ is the pupil function of the optical imaging system (i.e., a cylindrical step function with cutoff frequency $k_c = 2\pi NA/\lambda$ for a circular objective lens of numerical aperture NA [25]). With these prescriptions, the photocurrent signal produced by each pixel, which is proportional to $|E_{SPP}(\mathbf{r}_{sl}^n)|^2$, can be calculated for any given object as a function of pixel position $\mathbf{r}_{sl}^n$ across the array.

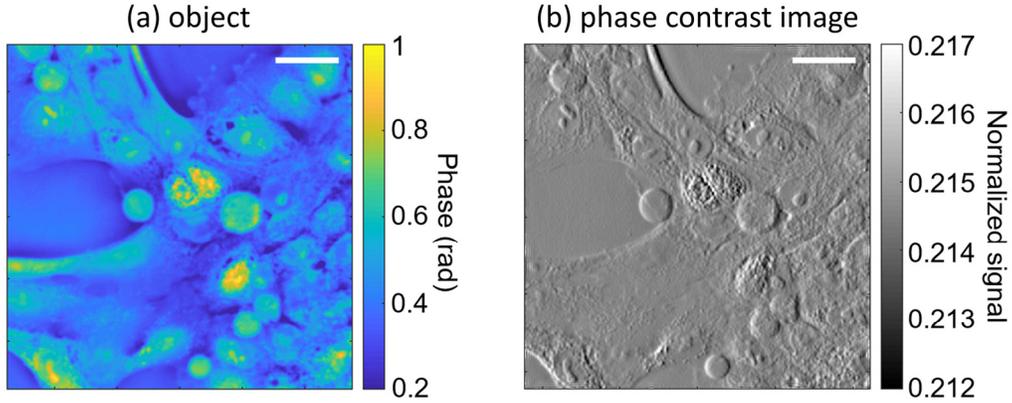

**Figure 3.** Computational phase contrast imaging results. (a) Representative phase object (MCF-10A cancer cells). (b) Corresponding image detected by an array of 512×512 angle-sensitive pixels modeled using the experimental data of Fig. 2(d). The signal intensity in this plot is normalized to the photocurrent produced by an otherwise identical device without any metasurface under the same illumination conditions. The scale bars (referenced to the object space in both panels) are 50 μm.

As an illustration, we consider the phase object shown in Fig. 3(a) (a sample of epithelial MCF-10A cancer cells, from ref. 26). Using the method just described, we compute the corresponding image recorded by a sensor array consisting of 512×512 square pixels described by the responsivity data $\mathcal{R}(\mathbf{k})$ of Fig. 2(d), combined with a telecentric 40× magnification system with NA = 0.8. Despite the transparent nature of the simulated object, a well resolved image is obtained, as shown in Fig. 3(b). Specifically, the detected signals at the cell edges are enhanced or decreased relative to the uniform background depending on the sign of the edge phase gradient along the horizontal (x) direction, in accordance with the asymmetric variation of $\mathcal{R}$ versus $k_x$ around normal incidence. The resulting image contrast is therefore maximum for vertically oriented edges, and steadily decreases for edges oriented towards the horizontal direction. This anisotropy is also found in other DPC techniques [4, 18, 19]. In the present approach, it could be eliminated by alternating pixels with orthogonally oriented nanostripes in a checkerboard pattern across the sensor array, as described in more details below and in the Supplementary Material. It should also be noted that the ~1550-nm operation wavelength of the present devices is not optimal for visualizing biological samples due to the background infrared water absorption. Nevertheless, the complex phase distribution of Fig. 3(a) provides a particularly vivid illustration of the phase-imaging capabilities of these devices. The extension of the same device concept to visible wavelengths and broadband operation is addressed in the conclusion section.

Next, we estimate the minimum detectable phase contrast with the metasurface of Fig. 2. For that purpose, we consider a simpler phase object consisting of y-oriented grating lines of variable contrast $\Delta\varphi$ [Fig. 4(a)]. The phase slope at the line edges is taken to be as large as possible, but small enough to avoid any noticeable pixelation in the detected image. Figure 4(b) shows the resulting photocurrent signal I(x) as a function of pixel position, computed with the same procedure above and normalized to the photocurrent of identical uncoated photodetectors under the same illumination conditions. Following ref. 19, the grating lines of Fig. 4(a) can be regarded as detectable if the contrast-to-noise ratio of the image $\text{CNR} = \frac{\Delta I}{I_{bg}} \text{SNR}(I_{bg})$ is larger than 1. Here, $\Delta I = |I_{\text{max(min)}} - I_{bg}|$ is the image contrast, where $I_{\text{max(min)}}$ is the maximum (minimum) signal at the positive (negative) edges of the grating lines, and $I_{bg}$ is the background signal away from the edges [see Fig. 4(b)]. The parameter $\text{SNR}(I_{bg})$ is the signal-to-noise ratio at the background signal level, which depends on the photodetector characteristics. For this analysis, we consider high-performance image-sensor photodiodes, where the dominant noise mechanism is generally shot noise and therefore the SNR is proportional to the square root of the signal. For optimized near-infrared photodiodes of comparable dimensions as the present devices, a SNR at full well capacity $SNR_{sat}$ = 71.3 dB (3,670×) can be achieved [27]. Furthermore, in the envisioned imaging system the optical source can be selected so that the pixels reach full capacity when illuminated at their angle of peak detection, where the photocurrent signal (again normalized to an identical uncoated device) is $I_{peak}$ = 30% [from Fig. 2(d)]. Based on all these considerations, $SNR(I_{bg})$ can be determined from the background signal $I_{bg}$ according to $SNR(I_{bg}) = SNR_{sat}\sqrt{I_{bg}/I_{peak}}$.

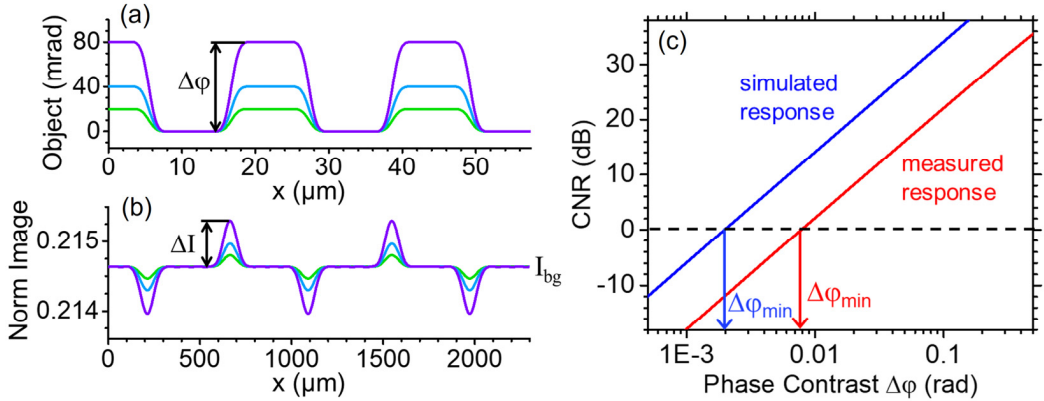

**Figure 4.** Minimum detectable phase contrast analysis. (a) Phase profiles of a one-dimensional grating for different values of the phase contrast Δφ. (b) Line cuts of the corresponding images detected by a 2D array of angle-sensitive pixels modeled using the experimental data of Fig. 2(d), combined with a 40× magnification system with NA = 0.8. The horizontal-axis labels in this plot refer to the pixel-array space. (c) Contrast-to-noise ratio (CNR) versus phase contrast Δφ for the object of (a), computed using the measured (red line) and calculated (blue line) angular response maps of the devices of Fig. 2. The vertical arrows indicate the minimum detectable values of Δφ, below which CNR < 1.

Figure 4(c) shows the CNR computed with this model as a function of the object phase contrast Δφ, with the image (and therefore $I_{max}$, $I_{min}$, and $I_{bg}$) evaluated using the measured responsivity map $\mathcal{R}(\mathbf{k})$ of Fig. 2(d) (red line) and the calculated map of Fig. 2(c) (blue line). As indicated by the arrows in the same plot, the minimum detectable phase contrasts obtained from these traces are 8 mrad and 2 mrad, respectively. These values are on par with the sensitivity limits of standard DPC techniques [19], which are based on more complex and bulkier setups as described above. Even smaller phase contrasts ($\lesssim$ 1 mrad) can be detected using interferometry [28] or a recently reported lock-in detection scheme [29], at the expense however of a further increase in system and measurement complexity. The results plotted in Fig. 4(c) therefore indicate that the present approach is fully suitable for high-sensitivity phase imaging applications, with the distinct advantage of enhanced miniaturization and portability. The comparison between the two traces in this figure also shows that, while the sensitivity is somewhat degraded by fabrication imperfections, state-of-the-art performance is still predicted for the experimental metasurfaces reported in this work, when combined with optimized image sensors.

Our devices also naturally lend themselves to single-shot quantitative phase reconstruction, using the array configuration shown schematically in Fig. 5(a). Here the array is partitioned into blocks of four adjacent pixels, each coated with the metasurface of Fig. 2 oriented along one of four orthogonal directions. In the following discussion, each type of pixels will be labeled by the unit vector perpendicular to the metasurface nanostripes and pointing away from the slits ($\hat{\mathbf{u}} = \pm\hat{\mathbf{x}}$ or $\pm\hat{\mathbf{y}}$). The photocurrent signals $I_{\hat{\mathbf{u}}}(\mathbf{r})$ measured by all pixels of each type across the whole array as a function of pixel-block center position $\mathbf{r}$ provide an edge-enhanced image of the phase object [such as for example Figs. 3(b) and 4(b) for $\hat{\mathbf{u}} = +\hat{\mathbf{x}}$, and Fig. S4(c) of the Supplementary Material for $\hat{\mathbf{u}} = -\hat{\mathbf{y}}$]. In these images, each edge of the phase object transverse to the $\hat{\mathbf{u}}$ direction produces a peak or a dip (depending on the sign of the edge slope) over a constant background, which in turn is proportional to the incident optical power P and thus is generally unknown.

Because of the asymmetric nature of these angle-sensitive devices, a peak over the background in $I_{+\hat{u}}(r)$ corresponds to a dip in $I_{-\hat{u}}(r)$ and vice versa. As a result, if the readout signals of the two pixels oriented along equal and opposite directions in each block are digitally normalized to their sum and subtracted from each other, the unknown background is subtracted out. The resulting signals

$$S_u(r) \equiv \frac{I_{+\hat{u}}(r) - I_{-\hat{u}}(r)}{I_{+\hat{u}}(r) + I_{-\hat{u}}(r)} \tag{2}$$

(for u = x and y) can therefore be used for quantitative phase reconstruction.

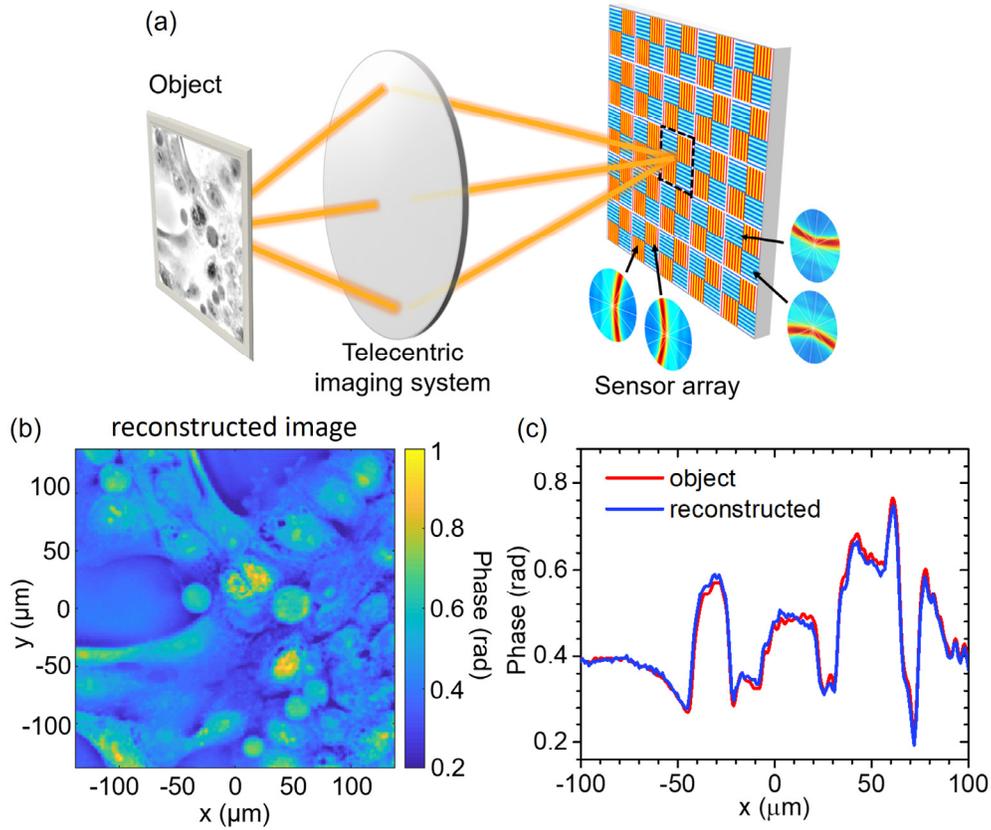

**Figure 5.** Computational quantitative phase imaging results. (a) Measurement protocol, where the sensor array is partitioned into blocks of four adjacent pixels coated with the metasurface of Fig. 2 oriented along four orthogonal directions. One representative pixel block is indicated by the dashed lines. The experimental angular response maps of all four pixels in each block are also shown. (b) Reconstructed phase distribution of the MCF-10A cell sample of Fig. 3(a). (c) Red trace: phase profile along the horizontal line at y = 0 of the same sample. Blue trace: reconstructed phase profile from (b).

In particular, for a pure phase object with sufficiently small phase $\varphi(\mathbf{r})$, the Fourier transforms of $S_u(\mathbf{r})$ and $\varphi(\mathbf{r})$ are linearly proportional to each other, i.e.,

$$S_u(\mathbf{k}) = H_u(\mathbf{k})\varphi(\mathbf{k}), \tag{3}$$

with transfer function (for $u = x$)

$$H_x(\mathbf{k}) = it_{\text{lens}}(\mathbf{k})\left\{\sqrt{\frac{\mathcal{R}(\mathbf{k})}{\mathcal{R}(0)}} - \sqrt{\frac{\mathcal{R}(-\mathbf{k})}{\mathcal{R}(0)}}\right\}, \tag{4}$$

where $i$ is the imaginary unit and $\mathcal{R}(\mathbf{k})$ is the responsivity map of Fig. 2. For $u = y$, the same expression applies with $\mathcal{R}(\mathbf{k})$ rotated by 90°. The key role played by the asymmetric nature of our devices is clearly evidenced in eq. (4), where the transfer function $H_u(\mathbf{k})$ would be identically zero for a symmetric responsivity map subject to $\mathcal{R}(\mathbf{k}) = \mathcal{R}(-\mathbf{k})$. The derivation of eqs. (3) and (4) is detailed in the Supplementary Material and builds on prior work on quantitative DPC imaging with time-modulated directional sources [18]. A similar expression can also be derived for the more general case of an object that introduces both amplitude and phase modulation upon light transmission or reflection. Importantly, the transfer function of eq. (4) does not depend on the incident optical power P and is fully determined by intrinsic properties of the imaging optics [$t_{\text{lens}}(\mathbf{k})$] and of the image sensors [$\mathcal{R}_u(\mathbf{k})$, which can be measured in the initial device calibration as in Fig. 2(d)]. Therefore, the phase profile $\varphi(\mathbf{r})$ can be retrieved quantitatively from the measured images $S_u(\mathbf{r})$ by inverting eq. (3). To avoid numerical artifacts associated with the zeros of the transfer function $H_u(\mathbf{k})$, we use the Tikhonov inversion method [18], whereby the reconstructed profile is

$$\varphi(\mathbf{r}) = \mathcal{F}^{-1}\left\{\frac{\sum_{u=x,y} H_u^*(\mathbf{k})S_u(\mathbf{k})}{\sum_{u=x,y}|H_u(\mathbf{k})|^2 + \alpha_T}\right\}. \tag{5}$$

In this equation, $\mathcal{F}^{-1}\{\}$ indicates the inverse Fourier transform, $\alpha_T$ is a regularization parameter, and both $S_x(\mathbf{r})$ and $S_y(\mathbf{r})$ are used simultaneously to allow for isotropic phase reconstruction.

An illustration of this protocol is shown in Figs. 5(b) and 5(c) for the phase object of Fig. 3(a). Here the edge-enhanced images recorded by the four types of pixels in the sensor array, i.e., $I_{\pm\hat{x}}(\mathbf{r})$ and $I_{\pm\hat{y}}(\mathbf{r})$, were computed with the frequency-domain model of eq. (1), again using the experimental data of Fig. 2(d) for the angle-dependent responsivity. Gaussian noise (with SNR estimated as described above) was then added to each image, and the results were used to evaluate the normalized signals $S_x(\mathbf{r})$ and $S_y(\mathbf{r})$ of eq. (2). In passing it should be noted that, in this normalization step, each peak and dip in the phase contrast image is automatically doubled in height, while the SNR is degraded by a factor of $\sqrt{2}$; as a result, the CNR is increased by $\sqrt{2}$, leading to a proportional decrease in the minimum detectable phase contrast. Given $S_x(\mathbf{r})$ and $S_y(\mathbf{r})$, eq. (5) was finally employed to reconstruct the phase profile $\varphi(\mathbf{r})$ of the MCF-10A-cell sample of Fig. 3(a). The result, shown in Fig. 5(b), is in excellent agreement with the original object. For a more direct quantitative comparison, the red and blue traces in Fig. 5(c) show, respectively, the original and reconstructed phase profile along the horizontal line at y = 0 of the same sample. Only very small discrepancies are observed in this plot, which are attributed to the weak-phase-object approximation used in the derivation of eqs. (3) and (4). A similar small-signal linear approximation is also used in standard DPC techniques for quantitative phase reconstruction, where multiple images of the object are recorded sequentially under different asymmetric illumination conditions [4, 18, 19]. The image sensors reported in this work thus allow for similar results, but with a significantly smaller system footprint and simpler measurement protocol. Furthermore, in the present approach, all the required images are collected simultaneously by the different types of pixels, which is beneficial for the purpose of increasing the frame rate (at the expense, however, of a proportional decrease in spatial resolution).

## 3. Conclusion

We have reported a new type of image sensors that allow for the direct visualization of transparent phase objects with a standard camera or microscope configuration. The key innovation of these devices is a metasurface coating that creates an asymmetric dependence of responsivity on illumination angle around normal incidence. This arrangement produces a high sensitivity to wavefront distortions caused by light propagation through a phase object, with state-of-the-art minimum detectable phase contrasts below 10 mrad. At the same time, the combination of pixels with equal and opposite angular response can be employed to normalize out the unknown incident power, and thus perform quantitative phase reconstruction in a single shot. The specific devices developed in the present work rely on a metallic metasurface design suitable for operation at near-infrared wavelengths, where plasmonic absorption losses are quite small. The same idea could also be extended to visible-range operation, e.g., by replacing SPPs with dielectric waveguide modes and the Au nanostripes with dielectric nanoparticles arranged in a gradient-metasurface architecture to introduce the required asymmetry. Similar configurations could also be designed to further tailor the angular response, including for example isotropic or vortex-like shapes, and to produce broadband achromatic operation by metasurface dispersion engineering. More broadly, our results also highlight a promising new research direction in flat optics, where metasurfaces are integrated directly within image sensor arrays to tailor their optical response on a pixel-by-pixel basis and correspondingly enable entirely new imaging capabilities.

## 4. Methods

**4.1 Design simulations.**

All the design simulations presented in this work were carried out with the Ansys-Lumerical FDTD Solutions software package. The angular response map in the inset of Fig. 2(c) was generated by computing the far-field radiation pattern in the air above the device for an electric dipole source positioned in the device substrate below the slits. In this simulation, a three-dimensional computational domain is employed, with perfectly matched layers (PMLs) on all boundaries. All relevant materials (Ge, $SiO_2$, Au) are described by their complex permittivity from a built-in database in the FDTD software. By reciprocity [30], the calculated pattern is proportional to the local field intensity at the dipole position produced by an incident plane wave as a function of illumination angles. This approach for computing the angular response of our devices is particularly convenient in terms of computational time, as all angles are covered in a single simulation. To calibrate the resulting color map, we have conducted additional two-dimensional simulations with Bloch boundary conditions on the lateral boundaries enclosing a full pixel. In these calculations, the metasurface is illuminated with a p-polarized plane wave and the transmitted light intensity into the device substrate is calculated for different values of the angle of incidence $\theta$ on the x-z plane. The results of these simulations [shown in the main plot of Fig. 2(c)] are qualitatively in good agreement with the horizontal line cut of the color map in the inset and allow calibrating its vertical axis to the metasurface transmission coefficient.

**4.2 Device fabrication.**

The experimental samples are fabricated on undoped (100) Ge substrates. The Au films (with a 5-nm Ti adhesion layer) are deposited by electron-beam evaporation, whereas RF sputtering is used for the $SiO_2$ layers. The slits are defined by electron-beam lithography (EBL) and reactive ion etching (RIE) with a positive/negative double layer of poly-methyl-methacrylate (PMMA) and hydrogen-silsesquioxane (HSQ) resist, followed by deposition of the Ti/Au film and liftoff. The Au nanostripes are patterned by EBL with a single positive resist (PMMA). The experimental samples consist of a few (7) identical repetitions of the structure of Fig. 2(a), with the reflector of one section immediately adjacent to the slits of the next section, and with a large (300 μm) separation between the two electrodes. This arrangement (equivalent to multiple identical pixels binned together) is convenient for the angle-resolved device characterization, because it alleviates the need for tightly focused incident light that would degrade the measurement angular resolution. In the final step of the fabrication process, a Ti window with an opening over the entire metasurface is deposited on the top $SiO_2$ layer and patterned by photolithography. This window is introduced to suppress any spurious photocurrent that may otherwise be caused by light absorbed near the electrodes away from the metasurface. The completed device is then mounted on a copper block and wire-bonded to two Au-coated ceramic plates.

### 4.3 Device characterization.

The measurement results presented in Fig. 2(d) were collected with a custom-built optical goniometer setup, where the device under study is biased with a 1-V dc voltage and illuminated with 0.5-mW linearly polarized light from a diode laser. The incident optical power is modulated at 1 kHz, so that the photocurrent can be measured separately from the dark current at low noise using a bias tee and lock-in amplifier. The laser light is delivered to the device with a polarization-maintaining fiber mounted in a cage system, which is rotated with a piezo-controlled stage about the focal point of its output lens to vary the polar angle of incidence $\theta$. The device is also mounted on another rotational stage that allows tuning the azimuthal illumination angle $\phi$. The polar angle is varied between ±85° in steps of 1°, whereas the measured azimuthal angles range from 0° to 90° in steps of 5°. The remaining two quadrants of the angular response maps are filled in based on the mirror symmetry of the device geometry. Finally, a linear interpolation is used to include additional data points between the measured values of $\phi$ in steps of 1°.

## Author Statements

### Corresponding author

Email: rpaiella@bu.edu

### Acknowledgements

The FDTD simulations were performed using the Shared Computing Cluster facility at Boston University.

### Research funding

This work was supported by the National Science Foundation under Grant # ECCS 2139451.

### Author contributions

All authors have accepted responsibility for the entire content of this manuscript and approved its submission.

### Conflict of interest

Authors state no conflicts of interest.

### Data availability

The datasets generated and/or analyzed during the current study are available from the corresponding author upon reasonable request.